\documentclass[preprint,12pt]{elsarticle}

\usepackage{amssymb}

\usepackage{color}
\usepackage{lineno}
\usepackage{url}
\usepackage{ulem}
\usepackage{lineno}

\journal{}

\begin{document}

\begin{frontmatter}



\title{Structural and Dynamical Changes in a Gd-Co Metallic Glass by Cryogenic Rejuvenation}


\author[KU]{Shinya~Hosokawa\corref{mycorrespondingauthor}}
\cortext[mycorrespondingauthor]{Corresponding author}
\ead{shhosokawa@kumamoto-u.ac.jp}
\author[NU]{Jens~R.~Stellhorn \fnref{cor1}}
\author[Wigner,IROAST]{L\'{a}szl\'{o}~Pusztai}
\author[IMR]{Yoshikatsu~Yamazaki \fnref {cor2}}
\author[IMR]{Jing~Jiang \fnref{cor3}}
\author[IMR]{Hidemi~Kato}
\author[IMR]{Tetsu~Ichitsubo}
\author[SagaLS]{Eisuke~Magome}
\author[CNRS]{Nils~Blanc}
\author[CNRS]{Nathalie~Boudet}
\author[JASRI]{Koji~Ohara \fnref {cor4}}
\author[JASRI,IU,IMR]{Satoshi~Tsutsui}
\author[JASRI]{Hiroshi~Uchiyama}
\author[RIKEN,JASRI]{Alfred~Q.~R.~Baron}

\address[KU]{Institute of Industrial Nanomaterials, Kumamoto University, Kumamoto, 860-8555, Japan}
\address[NU]{Department of Physics, Nagoya University, Nagoya, 464-8602, Japan}
\address[Wigner]{Wigner Research Centre for Physics, 1525, Budapest, Hungary}
\address[IROAST]{International Research Organization for Advanced Science and Technology (IROAST), Kumamoto University, Kumamoto, 860-8555, Japan}
\address[IMR]{Institute for Materials Research, Tohoku University, Sendai, 980-8577, Japan}
\address[SagaLS]{Kyushu Synchrotron Light Research Center, Tosu, 841-0005, Japan}
\address[CNRS]{Universit\'{e} Grenoble Alpes, CNRS, Grenoble INP, Institut N\'{e}el, 38000 Grenoble, France}
\address[JASRI]{Japan Synchrotron Radiation Institute (JASRI), Sayo, 679-5198, Japan}
\address[IU]{Institute of Quantum Beam Science, Graduate School of Science and Engineering, Ibaraki University, Hitachi, 316-8511, Japan}
\address[RIKEN]{Materials Dynamics Laboratory, RIKEN SPring-8 Center, RIKEN, Sayo, 679-5148, Japan}

\fntext[cor1]{Present address: Department of Materials Chemistry, Graduate School of Engineering, Kyoto University, Kyoto 615-8520, Japan}
\fntext[cor2]{Present address: Department of Mechanical Engineering, National Institute of Technology, Ube College, Ube 755-8555, Japan}
\fntext[cor3]{Present address: School of Health Sciences and Biomedical Engineering, Hebei University of Technology, Tianjin 300131, China}
\fntext[cor4]{Present address: Faculty of Materials for Energy, Shimane University, Matsue 690-8504, Japan}

\begin{abstract}
To experimentally clarify the changes in structural and dynamic heterogeneities in a metallic glass (MG), Gd$_{65}$Co$_{35}$, by rejuvenation with a temperature cycling (cryogenic rejuvenation), high-energy x-ray diffraction (HEXRD), anomalous x-ray scattering (AXS), and inelastic x-ray scattering (IXS) experiments were carried out. By a repeated temperature change between liquid N$_2$ and room temperatures 40 times, tiny but clear structural changes are observed by HEXRD even in the first neighboring range. Partial structural information obtained by AXS reveals that slight movements of the Gd and Co atoms occur in the first- and second-neighboring shells around the central Gd atom. The concentration inhomogeneity in the nm size drastically increases for the Gd atoms by the temperature cycling, while the other heterogeneities are negligible. A distinct change was detected in a microscopic elastic property by IXS: The width of longitudinal acoustic excitation broadens by about 20\%, indicating an increase of the elastic heterogeneity of this MG by the thermal treatments. These static and dynamic results explicitly clarify the features of the cryogenic rejuvenation effect experimentally.
\end{abstract}



\begin{keyword}

Metallic glasses \sep Cryogenic rejuvenation \sep Partial atomic structures \sep Phonon dynamics \sep Heterogeneity



\end{keyword}

\end{frontmatter}


\section{Introduction}
Rejuvenation in glasses is defined as an excitation to an higher energy state by an external stress, being the opposite of the usual relaxation process by thermal annealing. A rejuvenation effect by a temperature cycling in metallic glasses (MG) was recently reported by Ketov et al. on a La$_{55}$Ni$_{10}$Al$_{35}$ bulk MG \cite{Ketov}. According to their interpretation, the thermal expansion coefficient has a distribution over a glass sample if it is not elastically homogeneous. By repeated temperature changes, different magnitudes of thermal expansion at different positions in a glass induce shearing forces, and as a result, a rejuvenation effect occurs in the glass. They called this `Rejuvenation of metallic glasses by non-affine thermal strain' \cite{Ketov}. The validity of this picture is the subject of intensive debate.

Ketov et al. \cite{Ketov} found such a rejuvenation effect by temperature cycling between liquid N$_2$ and room temperatures (called `cryogenic rejuvenation' by Hufnagel \cite{Hufnagel}) in several macroscopic properties, including differential scanning calorimetry (DSC), instrumented indentation, compression test, resonant ultrasonic spectroscopy (RUS), and dynamic mechanical analysis (DMA). Concerning the microscopic structures, on the other hand, there is no discernible changes in x-ray diffraction (XRD) patterns as shown in Extended Data Fig. 6 of Ref. \cite{Ketov}, and they concluded that thermal cycling surely introduces heterogeneities that have small affect on the average atomic structure of MGs.

Hufnagel \cite{Hufnagel}, however, reviewed thermal cycling rejuvenation effect, and suggested that non-affine deformation must be caused on an atomistic length scale. He expects that simulations will reveal much about the fundamental atomic- and nanoscale mechanisms responsible for the rejuvenation. Note that Ketov et al.'s XRD data were taken by an in-house diffractometer in a limited scattering vector range of the first peak, corresponding to the intermediate-range of the MG structure. Thus, their conclusion may be premature. Detailed information about elementally selected (not averaged) atomic structures including the first-neighboring shell is highly desirable.

\begin{figure}
\begin{center}
\includegraphics[width=80mm]{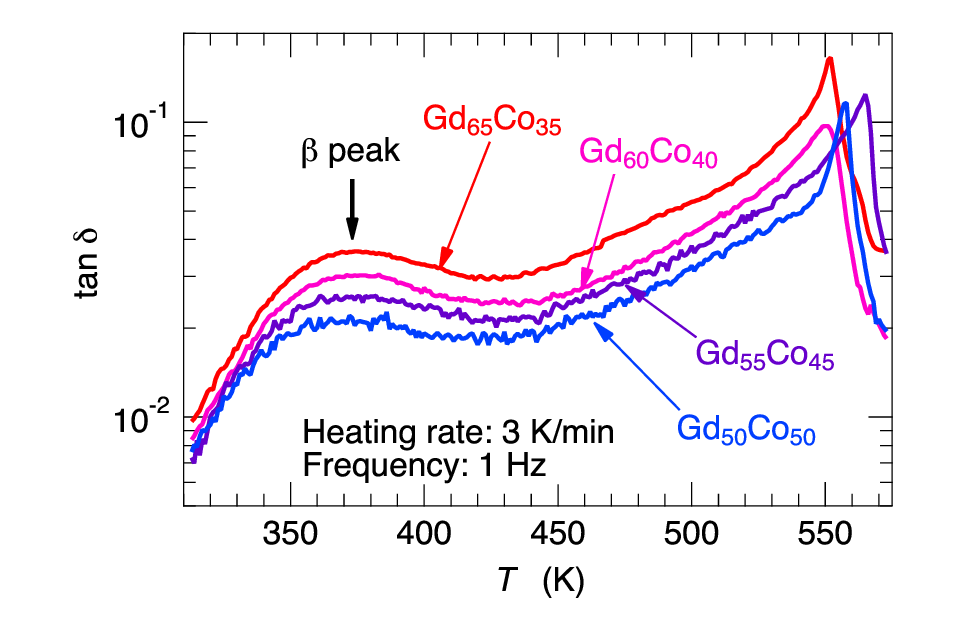}
\caption{\label{GdCobeta}Phase angles obtained from the DMA measurements on Gd$_{100-x}$Co$_x$ glasses \cite{Yamazaki}.}
\end{center}
\end{figure}

The extent of the heterogeneity of glasses can be judged by the so-called $\beta$-relaxation peak in DMA spectra \cite{Johari}, as argued by Ketov et al. \cite{Ketov}. Yamazaki measured DMA spectra on Gd$_{100-x}$Co$_x$ glasses \cite{Yamazaki}, and the obtained results of phase angle tan$\delta$ are shown in Fig. \ref{GdCobeta}. Distinct $\beta$-relaxation peaks were observed at about 370 K, and the largest peak was detected on the Gd$_{65}$Co$_{35}$ glass, where the largest heterogeneity was expected among the Gd-Co glass alloys. Thus, we chose the Gd$_{65}$Co$_{35}$ glass alloy as a sample for experimentally investigating the relationship between the rejuvenation effect and the structural and elastic heterogeneities of MG.

To find structural changes by the cryogenic rejuvenation, the overall features and the detailed element-selective knowledges were obtained by high energy XRD (HEXRD) and anomalous x-ray scattering (AXS), respectively. A direct method for investigating microscopic elastic properties is inelastic scattering. The elastic heterogeneity in a Pd-based MG was previously observed by inelastic x-ray scattering (IXS) \cite{IchitsuboPRB}. All the measurements were carried out before and after the temperature cycling between liquid N$_2$ and room temperatures 40 times. In this paper, we report structural and dynamical changes in a Gd$_{65}$Co$_{35}$ MG by the cryogenic rejuvenation. 

\section{Experimental procedure}
A master Gd$_{65}$Co$_{35}$ ingot was manufactured by arc-melting a mixture of pure Gd and Co metals in an Ar atmosphere. The purities of Gd and Co were 99.95 and 99.999 at.\%, respectively. Glassy foils with a thickness of about 20 $\mu$m and width of about 2 mm were prepared by melt spinning with a single Cu roll in a pure Ar atmosphere. The concentration was confirmed to be within 0.5 wt.\% of the nominal values by electron-probe micro-analysis. A thermal cycling treatment was made between liquid N$_2$ and room temperatures 40 times, and all the experiments were performed for the same sample foils before and after the above temperature cycling. 

The density of Gd$_{65}$Co$_{35}$ foil was measured using a pycnometer or constant volume method. For this, we used a Gay-Lussac-type glass pycnometer with a volume of about 25 cm$^3$ at 25$^\circ$C (Sansyo Co., Ltd. Type 82-2353). Liquid ethanol (FUJIFILM Wako Pure Chemical Corp., 99.5+\%) was used for the reference liquid with a density of $0.7849\times10^3$ kg/m$^3$ at 25$^\circ$C \cite{OIML}. A thermostat chamber (AS ONE Corp. Type FCI-280) was used for keeping the handling temperature at 25$^\circ$C during the density measurements. A dosage balance with a significant digits of 0.1 mg (AS ONE Corp. Type ITX-120) was used for measuring the weights. Since the sample has a small weight of some 100 mg with a small volume of some 10 mm$^3$, the error in the density mostly originates from the accuracy of weight.  The measurements were carried out totally 20 times, and a value of $(8.1\pm0.1)\times10^3$ kg/m$^3$ was obtained by removing 4 unusual results and averaging over the remaining 16 data. This value is similar to the averaged density of pure Gd and Co crystals. Note that no density change was detected beyond the experimental errors between those before and after the temperature cycling, which was also confirmed by the density estimation in the Fourier transform procedure of the HEXRD data.

High-energy XRD experiments were carried out at room temperature by using a diffractometer installed at BL04B2 \cite{Ohara} of the SPring-8, Sayo, Japan. The incident x-ray energy was set to be 112.83 keV. Diffraction signals from five foils with a total thickness of about 0.1 mm were measured by three pure Ge solid state detectors. The measurements were carried out in a wide $Q$ range up to 250 nm$^{-1}$, and Hanning functions were applied at both the edge of the $S(Q)$ experimental data to minimize the truncation errors in Fourier-transformed $g(r)$ functions.

\begin{figure}%
\centering
\includegraphics[width=80mm]{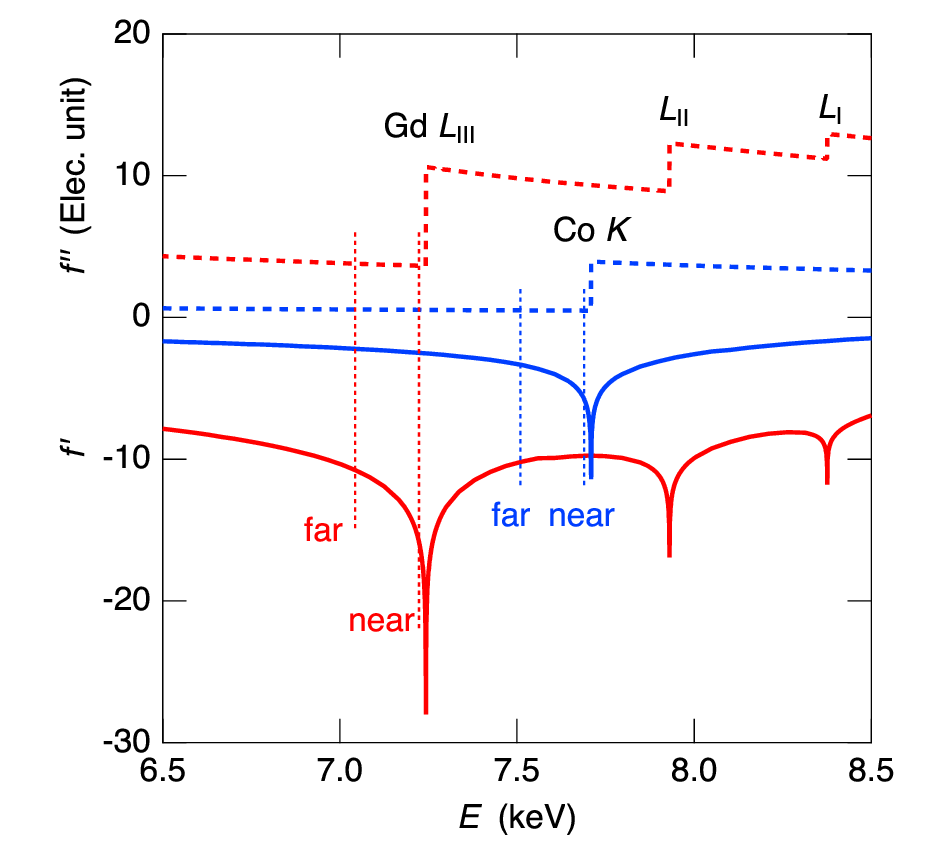}
\caption{\label{GdCof1f2}$f'$ (solid curves) and $f''$ (dashed curves) values of Gd (red) and Co (blue) in electron unit near the energies of the present AXS experiments indicated by vertical dotted lines. }
\end{figure}

AXS utilizes anomalous variations of atomic form factors, $f$, of a specific element near an x-ray absorption edge of the respective element \cite{Waseda}. The complex $f$ of an element is given as
$$f(Q,E)=f_0 (Q)+f'(E)+if"(E)$$
where $f_0$ is the usual energy-independent term, and $f'$ and $f"$ represent the real and imaginary parts of anomalous term, respectively. When the incident x-ray energy $E$ approaches an absorption edge of a constituent element, its $f'$ has a large negative minimum and $f"$ shows an abrupt jump as shown in Fig. \ref{GdCof1f2}. 

One can utilize the contrast between two scattering spectra near an absorption edge of the element $k$, $\Delta_kI$, where one is typically measured at some 10 eV and one at some 100 eV below the edge ($E^k_{\rm near}$ and $E^k_{\rm far}$, respectively). This differential intensity is connected to the differential structure factor $\Delta_kS(Q)$, expressed as
$$\alpha_k\Delta_kI(Q,E_{\rm near}^k, E_{\rm far}^k)=\Delta_k [\langle f^2\rangle-\langle f\rangle^2]+\Delta_k[\langle f\rangle^2 ]\Delta_k S(Q),$$
where $\alpha_k$ is a normalization constant and $\Delta_k [\hspace{2mm}]$ indicates the difference between the values in the bracket at energies of $E^k_{\rm far}$ and $E^k_{\rm near}$. The $\Delta_kS(Q)$ functions are given as a linear combination of partial structure factors $S_{ij}(Q)$ as
$$\Delta_kS(Q)=\sum_{i=1}^N\sum_{j=1}^NW_{ij}^k(Q,E_{\rm near}^k, E_{\rm far}^k)S_{ij}(Q)$$
Here, the weighting factors, $W_{ij}^k$, are given by
\begin{equation}
W_{ij}^k(Q,E_{\rm near}^k, E_{\rm far}^k)=x_i x_j\frac{\Delta_k[f_if_j ]}{\Delta_k[\langle f\rangle^2]},
\label{Wijk}
\end{equation}
where $x_i$ is the atomic concentration of element $i$. 

The AXS experiments were conducted using a standard $\omega$-2$\theta$ diffractometer installed at BL15 of the Saga Light Source (Saga-LS) in Tosu, Japan, and at the French Collaborating Research Group (CRG) BM02-D2AM of the European Synchrotron Radiation Facility (ESRF) in Grenoble, France. To obtain $\Delta_kS(Q)$s close to the Gd $L_{\rm III}$ (7.243 keV) and Co $K$ (7.709 keV) edges, two scattering experiments were carried out at energies 20 and 200 eV below the edges. Details of the experimental setups and data analyses are given elsewhere \cite{Stellhorn, HosokawaZPCGeSe, HosokawaGeSePRB, HosokawaZPCrev}. 
	
Figure \ref{GdCof1f2} shows the theoretical $f'$ (solid curves) and $f''$ (dashed curves) values of Gd (red) and Co (blue) in electron unit \cite{Sasaki} near the energies of the present AXS experiments indicated by vertical dotted lines. The $W_{ij}^k$ values were calculated by Eq. (\ref{Wijk}), and the obtained results at the $S(Q)$ maximum of $Q=23$ nm$^{-1}$ are listed together with those of $S(Q)$ in Table \ref{Wkij}. The bold numbers in the table indicate the large values with respect to those of $S(Q)$. Note that the $W_{\rm CoCo}^{\rm Co}$ value is 0.230, which is enough large for investigating the Co-Co correlation although the $x_{\rm Co}$ and $f_{\rm Co}$ values are quite small compared with those for Gd. It is also made aware that the $W_{\rm GdGd}^{\rm Co}$ value is a large negative value of -0.262, which is not usual in AXS, i.e., the Gd-Gd contribution is almost zero when the measured edge does not concern a Gd edge. In the case of this alloy, such an anomaly occurs because the Gd $L_{\rm III}$ and Co $K$ edges are near each other by less than 500 eV, and the $f'_{\rm Gd}$ value increases with increasing the x-ray energy from $E_{\rm far}^{\rm Co}$ to $E_{\rm near}^{\rm Co}$ as shown in Fig. S5 in the supplemental material, which induces a negative value  in $\Delta_{\rm Co}[f_{\rm Gd}f_{\rm Gd}]$ and accordingly a negative $W_{\rm GdGd}^{\rm Co}$ value as shown in Table \ref{Wkij}.

\begin{table}
\begin{center}
\caption{\label{Wkij}Weighting factors $W^k_{ij}$ at $Q=23$ nm$^{-1}$.}
 \begin{tabular}{crrr}
Data&Gd-Gd&Gd-Co&Co-Co\\
\hline
$S(Q)$&0.673&0.295&0.032\\
$\Delta_{\rm Gd}S(Q)$&{\bf 0.767}&0.225&0.008\\
$\Delta_{\rm Co}S(Q)$&-0.262&{\bf 1.032}&{\bf 0.230}\\
  \end{tabular}
\end{center}
\end{table}

RMC modeling \cite{McGreevy} is a useful tool to construct 3D structural models of disordered materials by using experimental diffraction data. In the RMC modeling, atoms of an initial configuration are moved so as to minimize the deviation from experimental structural data, e.g., in this study, two $\Delta_kS(Q)$s, $S(Q)$, and $g(r)$, with a standard Metropolis Monte Carlo algorithm \cite{Metropolis}. 

The starting configuration of a system containing 10,000 atoms with an estimated number density were generated by a hard-sphere Monte Carlo simulation, i.e., a random configuration with constraints of the proper shortest interatomic distances of 0.310, 0.255, and 0.255 nm for the Gd-Gd, Gd-Co, and Co-Co correlations, respectively. The RMC fits were carried out using RMC++ program package coded by Gereben et al. \cite{Gereben}. The RMC fits were started from 10 different initial conditions of the above-mentioned random configurations, and the obtained results were averaged over 10 different resultant atomic arrangements to improve the statistical quality of the results. 

To clarify the extent of the heterogeneity of the elemental concentrations quantitatively, the RMC simulation box was separated into $5\times5\times5=125$ voxels, and the number of each element were counted as was done for Pd-transition metal (TM)-P metallic glasses \cite{HosokawaPNP, HosokawaPCP, HosokawaPNCP}. 
To find the rejuvenation effect in the number density and concentration heterogeneities, fits were carried out by using Gauss functions. A similar analytical method was recently applied to clarify the size dependence of compositional fluctuations in Ti alloys \cite{Tane, Okamoto}. 

IXS experiments were carried out in transmission mode using a high energy-resolution IXS spectrometer \cite{Baron} installed at BL35XU of the SPring-8, Sayo, Japan. A monochromatized beam of approximately 10$^{10}$ photons/s was obtained from an undulator source through a cryogenically cooled Si(111) double crystal followed by a Si(11 11 11) monochromator operating in an extreme backscattering geometry (about 89.98$^\circ$ at 21.747 keV). The same backscattering geometry of twelve two-dimensionally curved Si analyzers on a 10 m detector arm was used for the energy analysis of the scattered x-ray photons. The IXS signals were collected near the sample positions with CdZnTe detectors. 
	
The energy scan was performed by changing the temperature of the monochromator by a few K, i.e., by utilizing the thermal expansion of the Si crystal, while the temperatures of the analyzer crystals were kept unchanged within some mK. The temperature/energy ratio is about 18 mK/meV. The energy resolutions were determined by the scattering from a Plexiglas sample, and fell within 1.5-1.8 meV full-width at half-maximum for various analyzer crystals. The $Q$ resolution was set to be about $\pm0.30$ nm$^{-1}$. 

To evaluate the excitation energies and widths of longitudinal acoustic (LA) phonon excitations, IXS spectra were analyzed by a damped harmonic oscillator (DHO) model \cite{Fak} expressed as 
\begin{equation}
\left[\frac{1}{1-e^{-\hbar\omega/k_BT}}\right]\frac{A_Q}{\pi}\frac{4\omega\omega_Q\Gamma_Q}{(\omega^2-\omega_Q^2 )^2+4\Gamma_Q^2\omega^2}
\label{DHO}
\end{equation}
Here, $A_Q$, $\omega_Q$, and $\Gamma_Q$ represent the amplitude, energy, and width close to half-width at half-maximum, of the excitation modes, respectively. A Lorentzian was assumed for the central quasi-elastic scattering. Fits to the experimental IXS data were carried out by the model functions convoluted with the experimentally obtained resolution functions. 
	
The DHO fit results frequently give large errors for the obtained parameters owing to parameter correlations. Since the fit parameters are assumed to gradually vary with $Q$, an idea of sparse modeling is introduced. The analytical technique is called `least absolute shrinkage and selection operator (LASSO)' proposed by Tibshirani \cite{Tibshirani}. In the fitting procedure of the sparse modeling, the error function $E$ is defined as,
\begin{equation}
E=\sum_i \mid S(Q,\omega_i)-\hat{S}(Q,\omega_i)\mid^2+\lambda\sum_j \mid \frac{p_j-p_{j0}}{p_{j0}}\mid, 
\nonumber
\end{equation}
where $S(Q,\omega_i)$ and $\hat{S}(Q,\omega_i)$ are the experimental and model functions at the experimental $\omega_i$ values, $p_j$ are the fit parameters in Eq. (\ref{DHO}), $p_{j0}$ are the smoothed values of $p_j$ by cubic function fits, and $\lambda$ is a penalty value of the fitting. For simplicity, a single $\lambda$ value was used. We started the fit results at $\lambda=0$, i.e., a usual least-squares fit, and the obtained parameters were smoothed over $Q$. Then, the fit to the data and the optimization of the $p_{j0}$ values were iterated by increasing $\lambda$ so as to keep the $E$ value not exceeding the original value by 3\%, corresponding to the error in the experimental data.

\section{Results}
\begin{figure}
\begin{center}
\includegraphics[width=80mm]{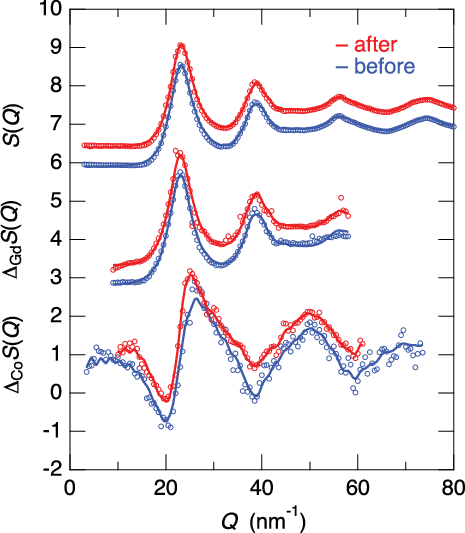}
\caption{\label{SQDSQ}From top to bottom, $S(Q)$ obtained from HEXRD, and $\Delta_{\rm Gd}S(Q)$ and $\Delta_{\rm Co}S(Q)$ from AXS. Circles and solid curves denote the experimental data and RMC fits, respectively. For clarity, spectra are displaced by 0.5 for before (blue) and after (red) the temperature cycling, and by 3 for spectra of different scattering processes. }
\end{center}
\end{figure}

Figure \ref{SQDSQ} shows, from top to bottom, the total structure factors, $S(Q)$, obtained from HEXRD, and the differential structure factors, $\Delta_{\rm Gd}S(Q)$ and $\Delta_{\rm Co}S(Q)$, from AXS, which are spectral contrasts near absorption edges and dominate the edge element-related partial structures. For clarity, spectra are displaced by 0.5 for before (blue) and after (red) the temperature cycling, and by 3 for spectra of different scattering processes. At a glance, there seems to be no differences between $S(Q)$s as well as $\Delta_{\rm Gd}S(Q)$s, while $\Delta_{\rm Co}S(Q)$s exhibit certain contrasts by the rejuvenation.

\begin{figure}
\begin{center}
\includegraphics[width=80mm]{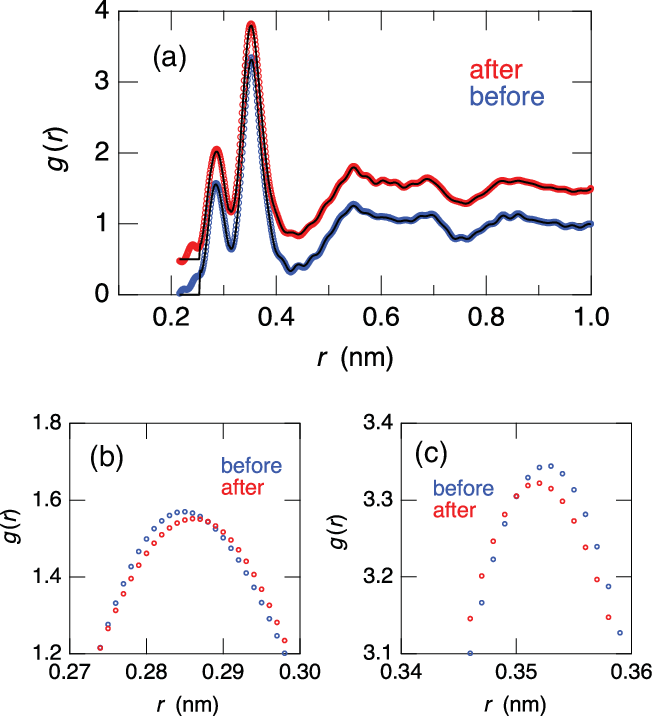}
\caption{\label{gr}$g(r)$s of Gd$_{65}$Co$_{35}$ MG before (blue) and after (red) the thermal cycling, where the latter is displaced upward by 0.5 for clarity. Zoom in on the (b) first- and (c) second peak regions, respectively (without the 0.5 offset). Circles and solid curves denote the HEXRD data and RMC fits, respectively.}
\end{center}
\end{figure}

Figure \ref{gr}(a) shows total pair distribution functions, $g(r)$, obtained from the Fourier transform of $S(Q)$. For this calculation, the number density of 39.707 nm$^{-3}$ was used for both the samples before and after the temperature cycling. In the first neighboring region, the $g(r)$ functions are separated into two peaks at about 0.284 and 0.352 nm. Hereafter, these peaks are called as the first- and second peaks, respectively. Figures \ref{gr}(b) and (c) indicate enlarged data in the first- and second peak regions, respectively. Both the peak heights become lower. The position of the first peak shifts toward the larger $r$ value by 0.011 nm, whereas that of the second peak moves toward the smaller $r$ value by -0.007 nm. Thus, structural differences by the rejuvenation can be obtained in the real space $g(r)$ data by Fourier-transforming $S(Q)$ in a wide $Q$ range up to more than 200 nm$^{-1}$. The HEXRD and AXS data were analyzed by reverse Monte Carlo (RMC) modeling. The quite good fit qualities are shown by solod curves in Fig. \ref{SQDSQ}(a) for $\Delta_kS(Q)$s obtained by AXS as well as $S(Q)$s by HEXRD and in Fig. \ref{SQDSQ}(b) for $g(r)$s. 

\begin{figure}
\begin{center}
\includegraphics[width=80mm]{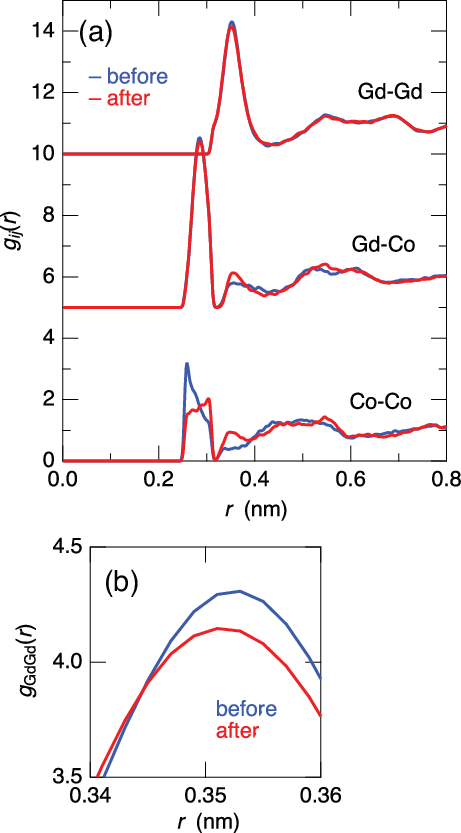}
\caption{\label{gijr} (a) From top to bottom, $g_{ij}(r)$s of Gd-Gd, Gd-Co, and Co-Co partial correlations obtained from the RMC analysis. The data are displaced upward by 5 for clarity. {\bf f.} (b) Zoom in on the peak of $g_{\rm GdGd}(r)$ in the upper panel (a).}
\end{center}
\end{figure}

Figure \ref{gijr}(a) shows partial pair distribution functions, $g_{ij}(r)$, of the Gd-Gd, Gd- Co, and Co-Co partial correlations from top to bottom obtained by the RMC fits before (blue) and after (red) the temperature cycling. Here, we begin discussing the changes in the partial structures by the cryogenic rejuvenation. Figure \ref{gijr}(b) shows the enlarged $g_{\rm GdGd}(r)$ data in the second neighboring peak region before (blue) and after (red) the temperature cycling. 

Here, we indicate the reliabilities of the partial functions by relating to the $W_{ij}^k$s given in Table \ref{Wijk} and the qualities of the experimental data with respect to the RMC fits shown in Figs. 1{\bf a} and 1{\bf b}. Since the $W_{\rm GdGd}^k$s in the $S(Q)$ and $\Delta_{\rm Gd}S(Q)$ are well beyond 2/3 and the RMC fits to these experimental data are excellent, the results of $g_{\rm GdGd}(r)$s are the most reliable partials. The second, and still very reliable partial is the Gd-Co correlation because the $W_{\rm GdCo}^k$s in the $S(Q)$ and $\Delta_{\rm Gd}S(Q)$ are about 30 and 20\%, respectively, and the qualities of the RMC fits are excellent. On the other hand, the Co-Co partial correlation is relatively unreliable since it is mainly determined experimentally by about 23\% of $\Delta_{\rm Co}S(Q)$. A plausible reason for the fit errors would be owing to the accuracy of the theoretical $f'$ values, which include an error of some \% near the absorption edge \cite{HosokawaGeSePRB}.

Although the spectral features are the same in the upper panel of Fig. \ref{gijr}(a), a clear difference is observed here in the peak region, i.e., by the cryogenic rejuvenation, the height of the main $g_{\rm GdGd}(r)$ maximum decreases and the peak position slightly shifts towards the lower $r$. These are features that have been observed in the second peak of total $g(r)$ as shown in Fig. \ref{gr}(c). The slight decrease of the peak height corresponds to the decrease of $N_{\rm GdGd}$ in the second neighboring region by about -0.05 given in Table \ref{Nij}.

Concerning the also reliable Gd-Co partials, the height of the first peak in $g_{\rm GdCo}(r)$ slightly decreases, and a small peak rises in the Gd-Co partial at the second peak positions by the temperature cycling, corresponding to a decrease and increase of $N_{\rm GdCo}$ and $N_{\rm CoGd}$ values slightly at these neighboring shells, respectively, as shown in Table \ref{Nij}. 

The Co-Co partial shows the distinct changes by the cryogenic rejuvenation, although the results are significantly less reliable compared with other partials as will be explained in the Materials and Methods: Reverse Monte Carlo modeling section later. The neighboring Co atoms around the central Co atom in the first peak region shift to either a longer $r$ in the first peak region or that in the second peak region as a peak.

\begin{figure}
\begin{center}
\includegraphics[width=80mm]{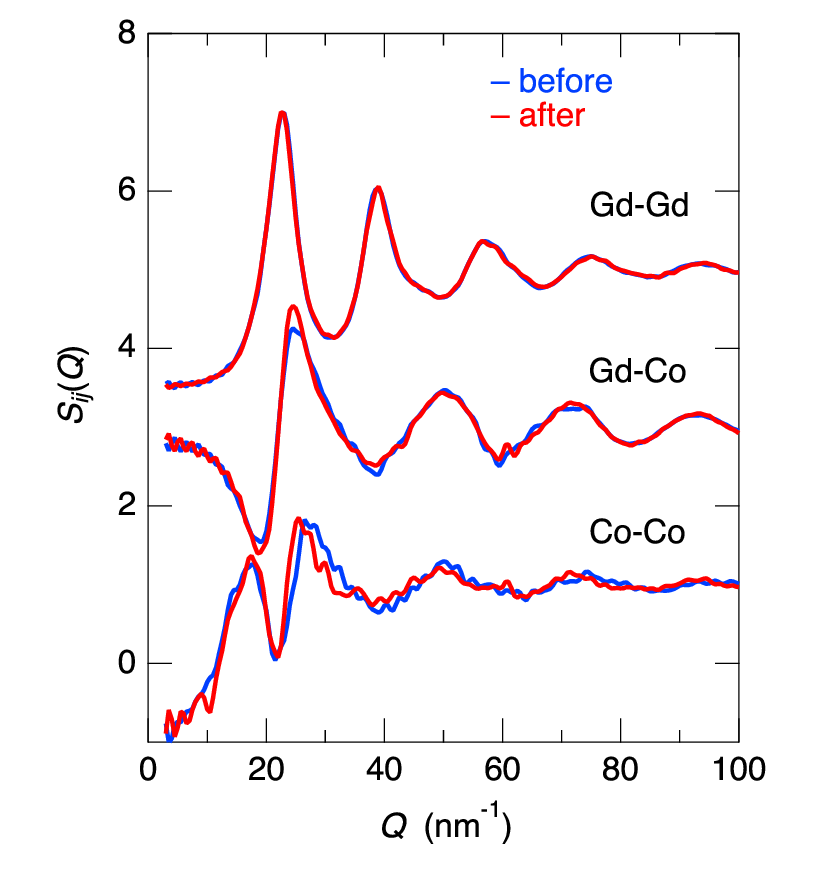}
\caption{\label{SijQ} The $S_{ij}(Q)$ spectra of the Gd-Gd, Gd-Co, and Co-Co partial correlations from top to bottom obtained from the RMC fits before (blue) and after (red) the temperature cycling treatment. For clarity, spectra are displaced by 2.}
\end{center}
\end{figure}

The corresponding partial structure factors, $S_{ij}(Q)$, are shown in Fig. S2 in the supplemental material. From the RMC fits, it is concluded that the first peak consists of the Gd-Co and Co-Co interatomic arrangements, while the second peak is mostly composed of the Gd-Gd interactions. Note that the Gd-Co interatomic distance is shorter than the average of the Gd-Gd and Co-Co lengths, indicating that this MG is not considered as a mixture of hard spheres of Gd and Co, but there is an additional correlation between them. Note that the obtained three $S_{ij}(Q)$s shown in Fig. \ref{SijQ} have rather reasonable features as a MG, although $\Delta_{\rm Co}S(Q)$ shown at the bottom of Fig. \ref{SQDSQ}(a) exhibits a quite complex spectral feature due to the abnormal $W_{ij}^{\rm Co}$s given in Table \ref{Wijk}. 

\begin{figure}
\begin{center}
\includegraphics[width=120mm]{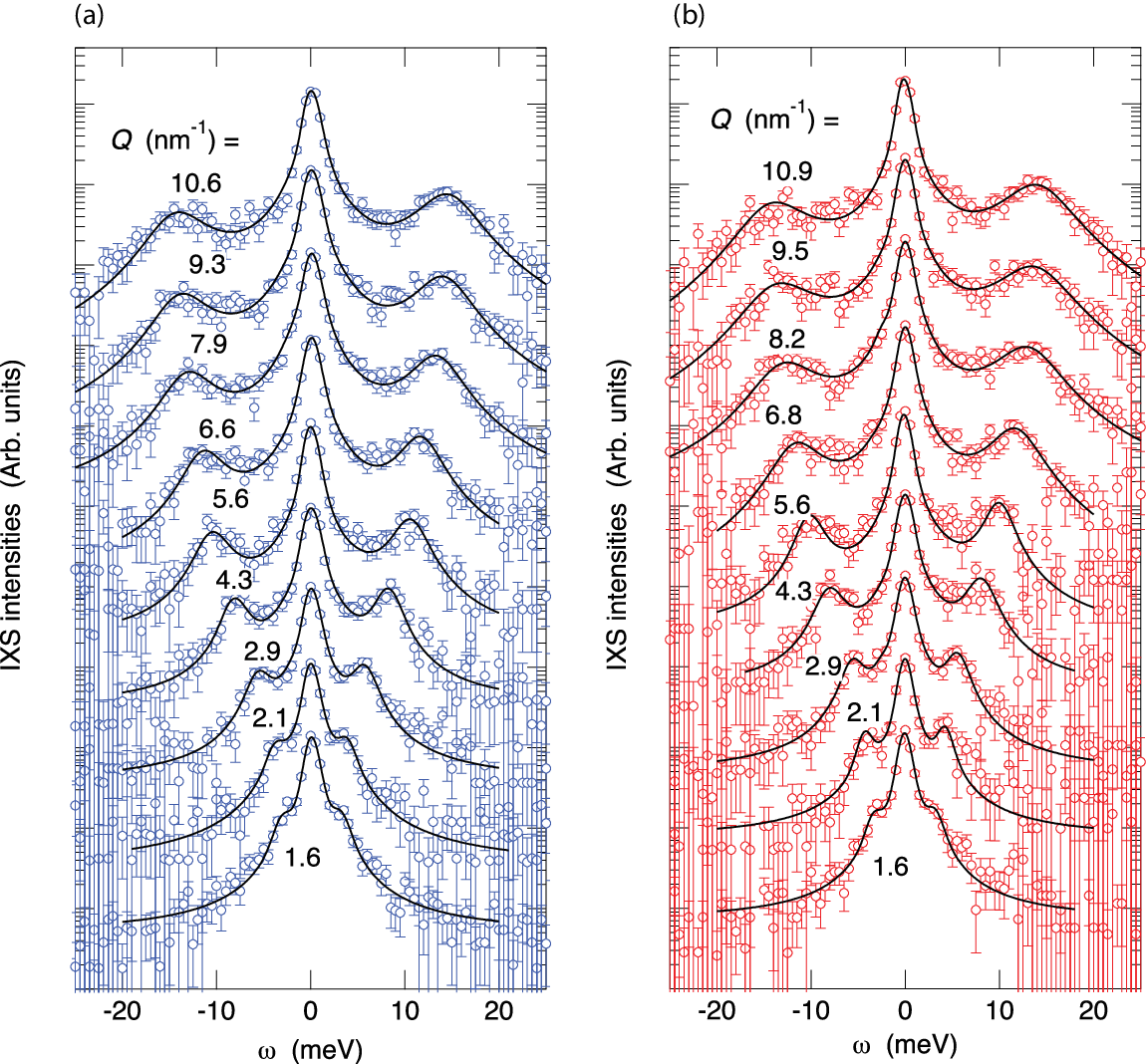}
\caption{\label{SQw} Logarithmic plots of IXS spectra of Gd$_{65}$Co$_{35}$ glass at selected $Q$ values (a) before and (b) after the temperature cycling. Solid curves indicate the fit results by a DHO model. For clarity, spectra are displaced by one order.}
\end{center}
\end{figure}

A direct method for investigating {\it microscopic elastic} properties is inelastic scattering. The elastic heterogeneity in MGs was observed by IXS \cite{IchitsuboPRB}. Figure \ref{SQw} shows logarithmic plots of IXS spectra of Gd$_{65}$Co$_{35}$ glass at selected $Q$ values (a) before and (b) after the temperature cycling. The central peaks indicate the quasi-elastic peaks, and the peaks or shoulders in the right- and left-hand sides represent respectively the Stokes and anti-Stokes signals of mainly longitudinal acoustic (LA) phonon excitation modes. 

By using a damped harmonic oscillator (DHO) model, the excitation energies, $\omega_Q$, and their widths, $\Gamma_Q$, can be evaluated for the LA phonon modes. Solid curves indicate the fit results by a DHO model. Good fit results are seen to the experimental data. 

\section{Discussion}
The partial coordination numbers of the $j$-th atoms around the $i$-th atoms, $N_{ij}$, are listed in Table \ref{Nij} before and after the cryogenic rejuvenation, and remarkable changes are shown as bold characters. The quantitative $N_{ij}$ values clearly confirm the above spectral features in $g_{ij}(r)$s. The sum of $N_{ij}$ around the Gd and Co atoms are approximately 12.7 and 10.0, the average of which is close to the densely packed value of 12. Note that these total coordination values slightly increase with the thermal treatment by 0.06 and 0.51 around Gd and Co, respectively.

\begin{table}
\begin{center}
\caption{$N_{ij}$ values around the first- and second neighboring shells obtained from the RMC fits before and after the temperature cycling. Remarkable changes by the cryogenic rejuvenation are given as bold characters.
}
\begin{tabular}{cc|cc}
$i$-$j$&shell&before&after\\
\hline
Gd-Gd&1st&0.3071(4)&0.3178(2)\\
&2nd&8.284(1)&8.230(1)\\
\hline
Gd-Co&1st&2.919(1)&2.892(1)\\
&2nd&1.110(1)&{\bf 1.245(2)}\\
\hline
Co-Gd&1st&5.422(1)&5.370(1)\\
&2nd&2.062(2)&{\bf 2.313(3)}\\
\hline
Co-Co&1st&1.559(2)&{\bf 1.405(2)}\\
&2nd&0.841(6)&{\bf 1.302(5)}\\
  \end{tabular}
  \label{Nij}
\end{center}
\end{table}

\begin{figure}
\begin{center}
\includegraphics[width=80mm]{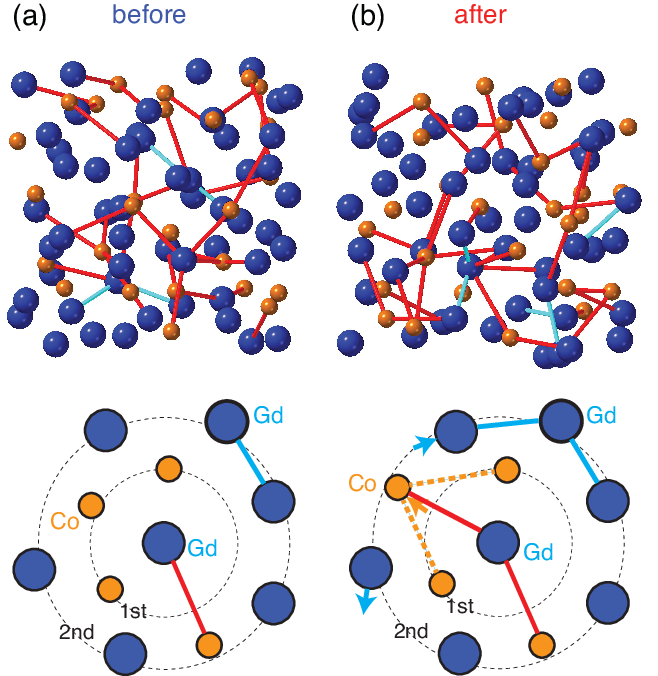}
\caption{\label{config} (Upper) typical atomic configurations obtained from the RMC fits for Gd$_{65}$Co$_{35}$ MG before and after the temperature cycling, respectively. (Lower) A structural model around the Gd atom for the rejuvenation effect.}
\end{center}
\end{figure}

The upper parts of Figs. \ref{config}(a) and (b) show typical atomic configurations obtained from the RMC fits before and after the temperature cycling, respectively. Large and small balls indicate the Gd and Co atoms, respectively. The blue and red lines between the atoms indicate the Gd-Gd bonds in the first neighboring distance and the Gd-Co bonds in the second neighboring distance, respectively. As seen in the figures, the numbers of these abnormal neighboring bonds increases by the thermal treatment. 

Based on the present partial structural data, schematic interpretations are given in the lower figures of Figs. \ref{config}(a) and (b) around the Gd atom. Before the temperature cycling, Co and Gd atoms are selectively located at the first and second neighboring shells, respectively, except a small number of bonds with different interatomic distances given by the lines, as shown in Fig. \ref{config}(a). By a temperature cycling treatment, Gd atoms move to either form a shorter Gd-Gd bonds or leave from the second peak region as shown by the blue arrows. The Gd-Co interatomic distance increases from the first to second neighboring distances as shown by the orange arrow. New longer Co-Co bonds may be created by this atomic movement as shown by dashed orange lines in the figure.

\begin{figure}
\begin{center}
\includegraphics[width=80mm]{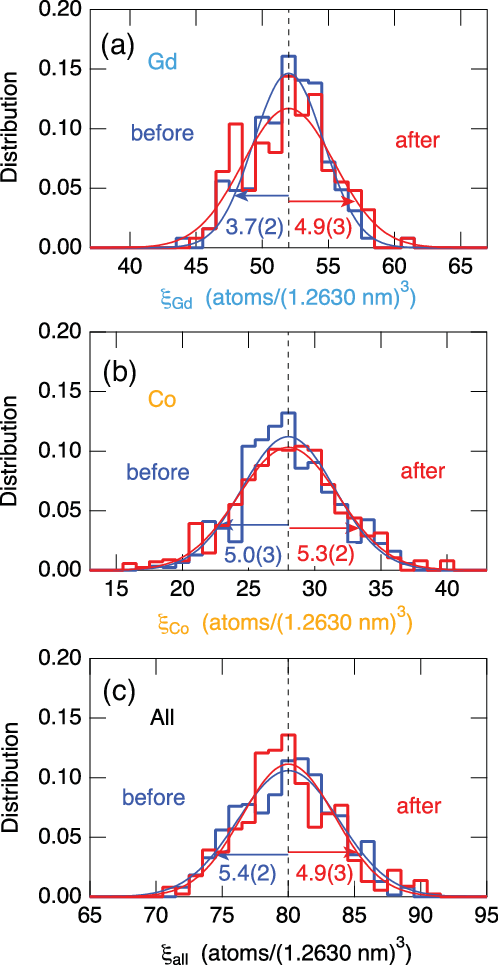}
\caption{\label{heterogeneity} Atomic number distributions of Gd, Co, and all atoms, respectively, in Gd$_{65}$Co$_{35}$ MG before (blue) and after (red) the temperature cycling.}
\end{center}
\end{figure}

The existence of heterogeneities in a MG was experimentally observed by Ichitsubo et al. \cite{IchitsuboPRL} for the first time by using an ultrasonic accelerated crystallization. Here, we examine the number density and concentration inhomogeneities by using the present RMC results. Figures \ref{heterogeneity}(a), (b), and (c) show the atomic number distributions of Gd, Co, and all the elements, respectively, in totally 125 voxels with the size of 1.2630 nm cubic, $\xi_i$ [atoms/(1.2630 nm)$^3$]. The averages are shown by the dashed lines in the figures; 52, 28, and 80 atoms for Gd, Co, and all the elements, respectively. The Gaussian fits are shown by thin solid curves, and the standard deviations, $\sigma_i$, are given in the figures before (blue) and after (red) the temperature cycling rejuvenation. The uncertainties for all of the $\sigma_i$ values, $\Delta\sigma_i$, are indicated in the parentheses. 

It should be noted that in the nm size length scale, the $\sigma_i$ values of Gd shown in Fig. \ref{heterogeneity}(a) increase by the rejuvenation of the Gd$_{65}$Co$_{35}$ MG, indicating that the concentration heterogeneity for the Gd atoms increases by the thermal cycling. On the other hand, the $\sigma_i$ values of Co and all elements, i.e., the number density heterogeneity, seem not to change beyond the errors by the thermal treatment.

\begin{figure}
\begin{center}
\includegraphics[width=80mm]{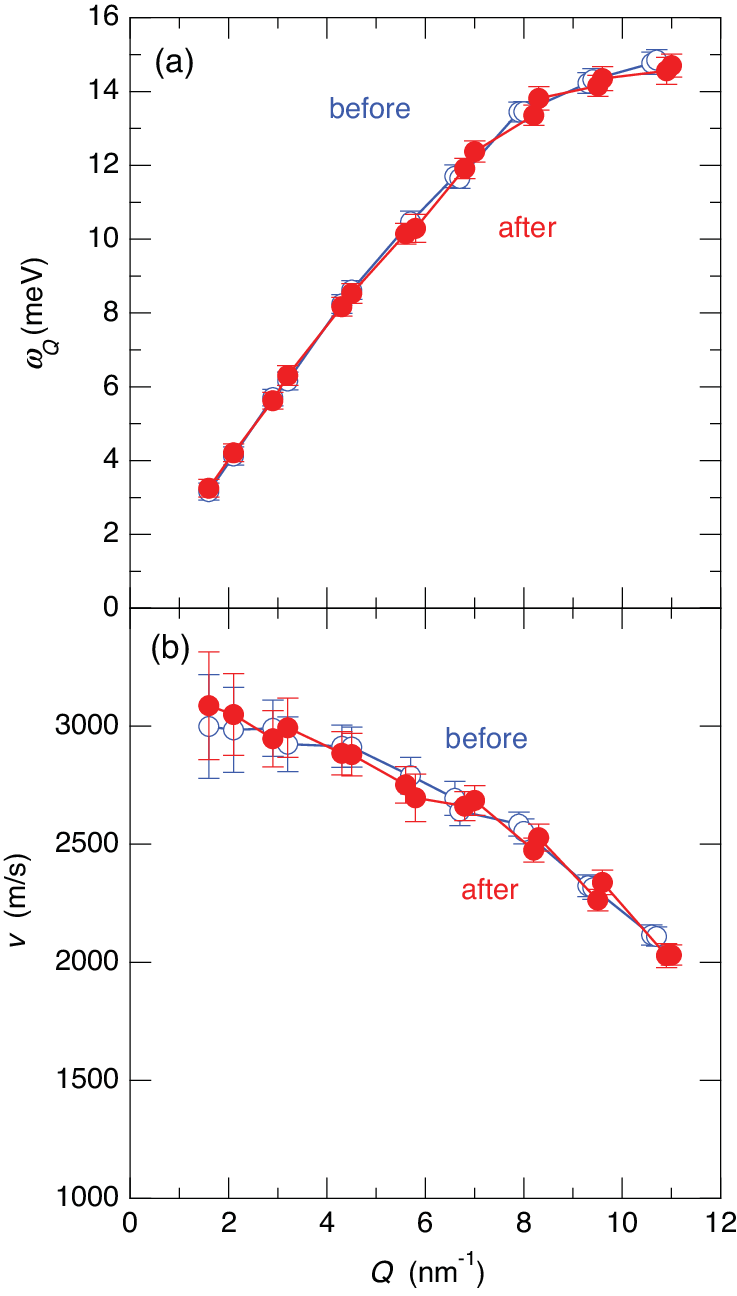}
\caption{\label{wQcQ} (a) Dispersion relations and (b) dynamical sound velocity $v$ of the LA excitation modes before (blue) and after (red) the temperature cycling.}
\end{center}
\end{figure}

Figure \ref{wQcQ} shows (a) the dispersion relations ($Q-\omega_Q$) of the LA phonon modes obtained from the DHO fits (a) before and (b) after the temperature cycling, and (b) the $Q$ dependence of dynamical sound velocity, $v=\omega_ Q/Q$, of the LA excitations. As seen in both the figure, no changes are observed within the experimental errors from the temperature cycling. The $Q\rightarrow0$ limit of high frequency $v$ is estimated to be $3100\pm100$ m/s before and after the thermal treatment. The relationship between $v$ and elastic moduli in glasses is given as $\rho v^2=K+\frac{4}{3}G$, where $K$ and $G$ are the bulk and shear moduli, respectively , and $\rho$ is the density. Therefore, the average value of the sum of elastic moduli in this glass seems to be unchanged to a precision of $\pm6$\% by the thermal rejuvenation.

\begin{figure}
\begin{center}
\includegraphics[width=80mm]{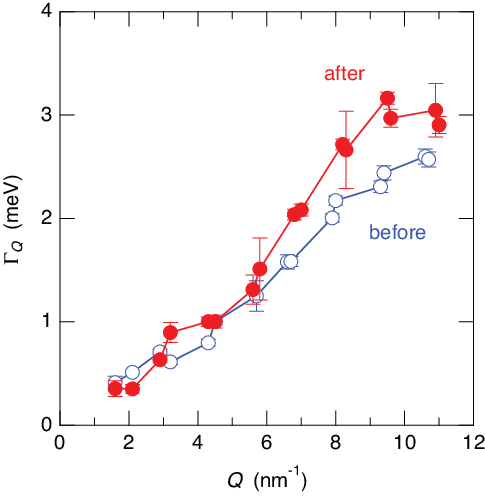}
\caption{\label{GammaQ} $Q$ dependence of width $\Gamma_Q$ of the LA excitations obtained from the DHO fit to the IXS spectra before (blue) and after (red) the temperature cycling.}
\end{center}
\end{figure}

Figure \ref{GammaQ} shows the $Q$ dependence of $\Gamma_Q$ values obtained by the DHO fits. In contrast with the $\omega_Q$ results, the $\Gamma_Q$ values strongly increase by about 20\% at higher $Q$ values after the temperature cycling. The sound attenuation indicated by the $\Gamma_Q$ values is modeled as the sum of the anharmonic (thermally dominant) and relaxational damping of terahertz LA waves \cite{Baldi}. The latter can be calculated \cite{Kawahara} and is proportional to the variance of the spatial fluctuations of the local sound velocity. This indicates that the present thermal treatment increases the distribution of local (or microscopic) elastic constants. In other words, the elastic heterogeneity increases by the cryogenic rejuvenation. Since the increase appears beyond about 6 nm$^{-1}$, the length scale of the inhomogeneity is about 1 nm. As seen in Figs. \ref{config}(a) and (b), a small and local free volume is expected to generate around the Gd atom, which may highly influence the elastic properties of this MG by the thermal treatment. 

\section{Summary}
In conclusion, structural and dynamical changes in Gd$_{65}$Co$_{35}$ MG by the rejuvenation with temperature cycling were experimentally investigated by HEXRD/AXS with RMC and IXS, respectively. The structural changes were mainly observed in the Gd-Gd and Gd-Co partial structures, where the Gd atoms move slightly in the second peak region and the Co atoms move from the first- to second-neighboring shells. The changes in the concentration inhomogeneity by the cryogenic rejuvenation were clearly found mainly for the Gd atoms in the nm size scale. From the elastic inhomogeneity studied by IXS clarified that the distribution of the elastic properties in Gd$_{65}$Co$_{35}$ MG becomes more extended by the thermal treatment although the average value of the sound velocity mostly unaffected. The cryogenic rejuvenation effect predicted by Ketov et al. \cite{Ketov} and expected to occur in the atomistic scale by Hufnagel \cite{Hufnagel}, was clearly realized by the present investigations for atomic structures as well as dynamic properties in Gd$_{65}$Co$_{35}$ MG experimentally.	Moreover, electronic states should be affected by the structural changes by the cryogenic rejuvenation, and the changes in some electron spectra will be exhibited later.

\section*{Acknowledgments}
HEXRD and IXS experiments were performed at BL04B2 (No. 2020A1500) and BL35XU (Nos. 2021A1292 and 2021B1111) of the SPring-8, respectively. AXS experiments were conducted at BL15 of the Saga-LS (No. 1902010A), and at BM02 of the ESRF (No. HC3438). This work was performed under the GIMRT Program of the Institute for Materials Research, Tohoku University (Nos. 202012-CRKEQ-0014 and 202012-RDKGE-0023). These works were supported by the Japan Society for the Promotion of Science (JSPS) Grant-in-Aid for Transformative Research Areas (A) `Hyper-Ordered Structures Science' (Nos. 21H05569 and 23H04117), that for Scientific Research (C) (No. 22K12662), and the Japan Science and Technology Agency (JST) CREST (No. JP-MJCR1861).




\section*{CRediT authorship contribution statement}
{\bf Shinya~Hosokawa} : Conceptualization, Methodology, Formal analysis, Investigation, Data curation, Writing - original draft, Writing - review \& editing, Visualization, Supervision, Project administration, Funding acquisition, {\bf Jens~R.~Stellhorn} : Formal analysis, Data curation, Investigation, {\bf L\'{a}szl\'{o}~Pusztai} : Formal analysis, {\bf Yoshikatsu~Yamazaki} : Investigation, Resources, {\bf Jing~Jiang} : Resources, {\bf Hidemi~Kato} : Resources, Conceptualization {\bf Tetsu~Ichitsubo} : Conceptualization {\bf Eisuke~Magome} : Methodology, Data curation, {\bf Nils~Blanc} : Data curation,  {\bf Nathalie~Boudet} : Methodology, Data curation,  {\bf Koji~Ohara} :  Methodology, Data curation, {\bf Satoshi~Tsutsui} :  Data curation, {\bf Hiroshi~Uchiyama} :  Data curation, {\bf Alfred~Q.~R.~Baron} : Methodology, Data curation, Writing - review \& editing. 

\section*{Declaration of Competing Interest}
The authors declare that they have no known competing financial interests or personal relationships that could have appeared to influence the work reported in this paper.

\section*{Competing interests}
The authors declare no competing interests.

\section*{Data availability}
The data that support the findings of this study are available from the corresponding author upon reasonable request.

\end{document}